\documentclass[journal,11pt]{IEEEtran}

\IEEEoverridecommandlockouts

\usepackage[nospace,noadjust]{cite}
\usepackage[english]{babel}
\usepackage[T1]{fontenc}
\usepackage{multirow}
\usepackage{amsmath,amssymb,amsfonts}
\usepackage{multicol}
\usepackage{multirow}
\usepackage{graphicx}
\usepackage{subfigure}
\usepackage{color}
\usepackage{textcomp}
\usepackage{soul}
\usepackage{dsfont}
\usepackage{tagging}
\usepackage{mathtools}
\graphicspath{{./images/}}

\usepackage{array}
\newcolumntype{L}{>{\centering\arraybackslash}m{3cm}}

\sethlcolor{yellow}
\usepackage[colorinlistoftodos,prependcaption,textsize=normalsize]{todonotes}
\usepackage{xargs}

\newcommandx{\doubt}[2][1=]{\todo[linecolor=red,backgroundcolor=red!25,bordercolor=red,#1]{#2}}

\usepackage[colorlinks=true, allcolors=blue]{hyperref}
\def\BibTeX{{\rm B\kern-.05em{\sc i\kern-.025em b}\kern-.08em
        T\kern-.1667em\lower.7ex\hbox{E}\kern-.125emX}}
\newcommand*{\email}[1]{%
    \normalsize\href{mailto:#1}{#1}\par
    }

\begin{document}
    
    \title{Arbitrary Scale Super-Resolution Assisted Lunar Crater Detection in Satellite Images
    }

    \author{
    \IEEEauthorblockN{Atal Tewari\IEEEauthorrefmark{2}, Nitin Khanna\IEEEauthorrefmark{3}$^{,\star}   
    $\thanks{$^{\star}$ Corresponding author. Please address all correspondences to Nitin Khanna, Multimedia Analysis and Security (MANAS) Lab, Electrical Engineering, Indian Institute of Technology Bhilai, India. E-mail address: \email{nitin@iitbhilai.ac.in}}
    }\\
    \IEEEauthorblockA{\IEEEauthorrefmark{2}Electrical Engineering, Indian Institute of Technology Gandhinagar, India}\\
    \IEEEauthorblockA{\IEEEauthorrefmark{3} Electrical Engineering, Indian Institute of Technology Bhilai, India}
}

\maketitle  
\thispagestyle{plain}
\pagestyle{plain}

\begin{abstract}

Craters are one of the most studied planetary features used for different scientific analyses, such as estimation of surface age and surface processes. 
Satellite images utilized for crater detection often have low resolution (LR) due to hardware constraints and transmission time. 
Super-resolution (SR) is a practical and cost-effective solution; however, most SR approaches work on fixed integer scale factors, i.e., a single model can generate images of a specific resolution.
In practical applications, SR on multiple scales provides various levels of detail, but training for each scale is resource-intensive.
Therefore, this paper proposes a system for crater detection assisted with an arbitrary scale super-resolution approach (i.e., a single model can be used for multiple scale factors) for the lunar surface. 
Our work is composed of two sub-systems. The first sub-system employs an arbitrary scale SR approach to generate super-resolved images of multiple resolutions.
Subsequently, the second sub-system passes super-resolved images of multiple resolutions to a deep learning-based crater detection framework for identifying craters on the lunar surface.
Employed arbitrary scale SR approach is based on a combination of convolution and transformer modules. 
For the crater detection sub-system, we utilize the Mask-RCNN framework.
Using SR images of multiple resolutions, the proposed system detects $13.47$\% more craters from the ground truth than the craters detected using only the LR images.
Further, in complex crater settings, specifically in overlapping and degraded craters, $11.84$\% and $15.01$\% more craters are detected as compared to the crater detection networks using only the LR images. 
The proposed system also leads to better localization performance, $3.19$\% IoU increment compared to the LR images.

\end{abstract}


\section{Introduction}
Craters are topographic features formed due to the impact of meteorites, volcanic activity, or an explosion~\cite{bandeira2012detection}. 
The study of the craters' characteristics, such as frequency, shape, and size are used to gain insight into the geological processes of planetary bodies without having to land on planetary surfaces~\cite{barlow2005review}.
For instance, crater density is associated with surface age, and crater counts have been utilized to estimate the relative age of planetary surfaces~\cite{michael2010planetary}. 
Crater detection can be addressed either via manual marking or by applying various image processing and deep learning techniques on satellite images. 
In the manual marking, domain experts visually inspect the planetary surface for marking craters~\cite{robbins2019new}. 
However, a manual process is laborious, time-consuming, and susceptible to errors. For example, Robbins et al.~\cite{robbins2014variability} reported that among experts, crater sizes and distribution might differ as much as $45\%$. 
Therefore, most studies have focused on automatic crater detection  approaches~\cite{silburt2019lunar,TEWARI2022105500,delatte2019segmentation,lee2019automated,lee2021automated,yang2020lunar,yang2022progressive}.

\begin{figure}
    \centering
    \includegraphics[width=0.49\textwidth]{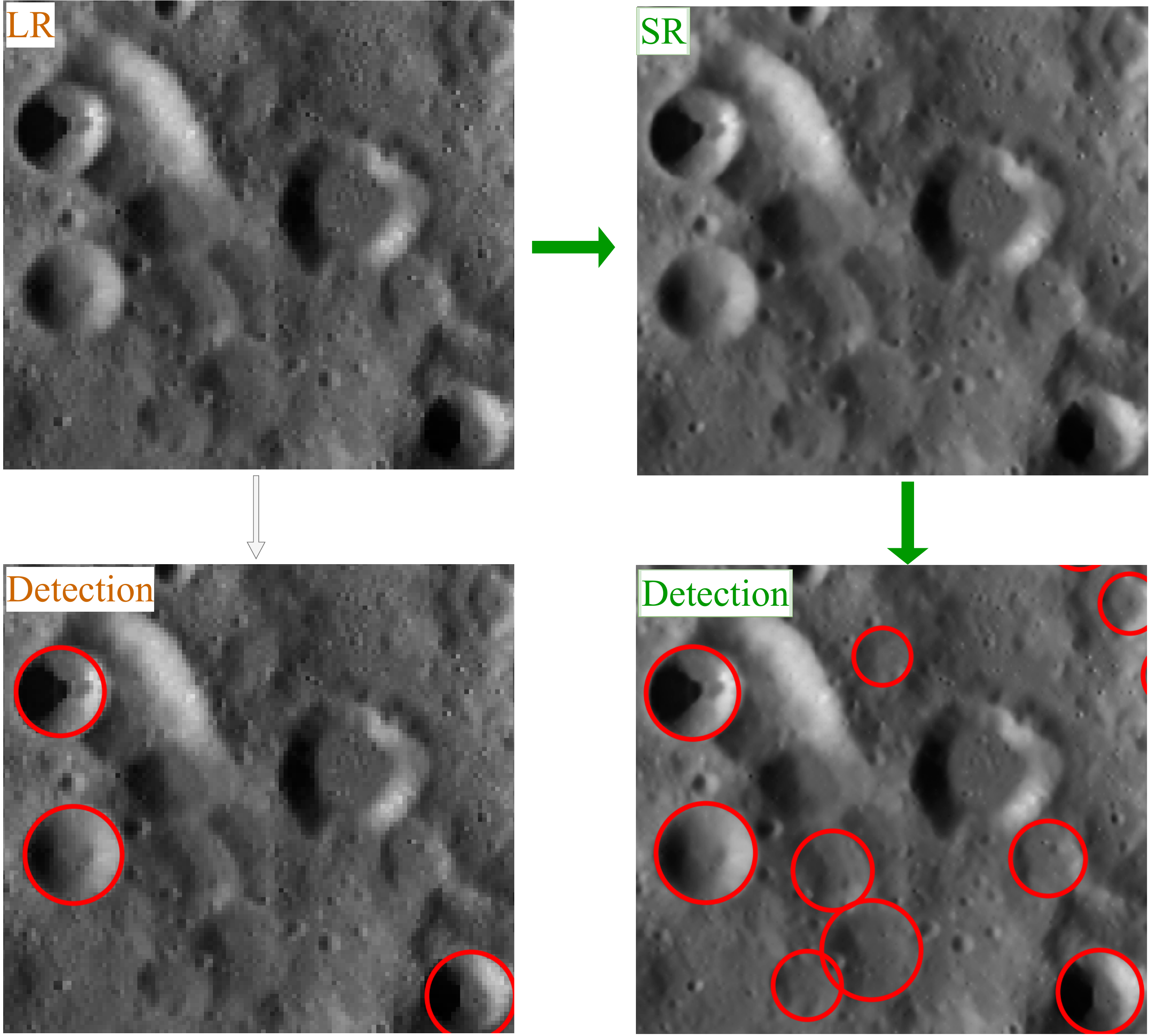}
    \caption{Crater detection performance on an SR-assisted image (proposed process shown by green arrow leads to higher number of detected craters).}
    \label{fig:intro_overview}
\end{figure}

The spatial resolution of images is one of the key factors for superior performance in crater detection. Higher-resolution images can contain more distinguishable features, which can help discriminate between crater and non-crater features. 
However, generating, transmitting and storing higher-resolution images involve various challenges such as need of high-quality sensors which are expensive. Therefore, super-resolution (SR) might be more practical and cost-effective in improving image quality. 
It will reduce the launch cost, decrease the number of sensors, and improve downlink speed~\cite{shermeyer2019effects}. 
Most of the existing super-resolution algorithms are based on fixed scale factors~\cite{Kim_2016_CVPR1, Tai_2017_CVPR,zhang2018image,tewari2021orbit,lu2019satellite}. 
Training the model for each scale separately might be impractical due to limited storage and computation power.
To address this problem, arbitrary scale super-resolution, which is capable of predicting high-resolution images in multiple-scale and non-integer scales from a single model, has gained attention.

Some existing works have applied super-resolution algorithms for better object detection performance~\cite{shermeyer2019effects,bilgazyev2011sparse}. 
However, it is mostly under-explored for thin atmosphere planetary surfaces such as Lunar and Mars. 
Super resolved images can be beneficial for detecting degraded, overlapping, and small craters, which are barely visible in low-resolution images. 
Therefore, we utilize an arbitrary scale super-resolution model to predict high-resolution images and assess it to understand the impact on crater detection performance.
The process of the system is shown in Figure~\ref{fig:intro_overview}. 
Following are the major contributions of this paper: 

\begin{itemize}
    \item  Presenting a solution for lunar crater detection using Swin Transformer-based arbitrary scale super-resolution.  

    \item Employing super-resolution images of multiple resolutions, we have achieved a $13.47$\% improvement in crater recovery from ground truth, as opposed to systems using only low-resolution images.
    
    \item In complex crater characteristics scenarios, i.e., overlapping and degraded craters, the increment of crater detection from ground truth is 11.84\% and 15.01\%, respectively, when evaluating predicted images from SR models of multiple resolutions, compared to LR images.  

    \item The IoU percentage is calculated to evaluate the localization performance; the combined outputs of SR-predicted images have a 3.19\% IoU increment compared to the LR images.

    \item Finally, by utilizing arbitrary scale super-resolution, we also show the effectiveness of the proposed system in detecting craters much smaller than training size without any retraining.

\end{itemize}

The rest of this paper is organized as follows. 
Section~\ref{sec:related_works} describes existing literature in the field.
The proposed system and its components are detailed in Section~\ref{sec:methods}. 
Further, Section~\ref{sec:experimental results} presents experimental results and their detailed analysis. 
Conclusions and future work are presented in Section~\ref{sec:Conclusion and Future Work}.

\begin{figure*}[ht!]
    \centering
    \includegraphics[width=\textwidth]{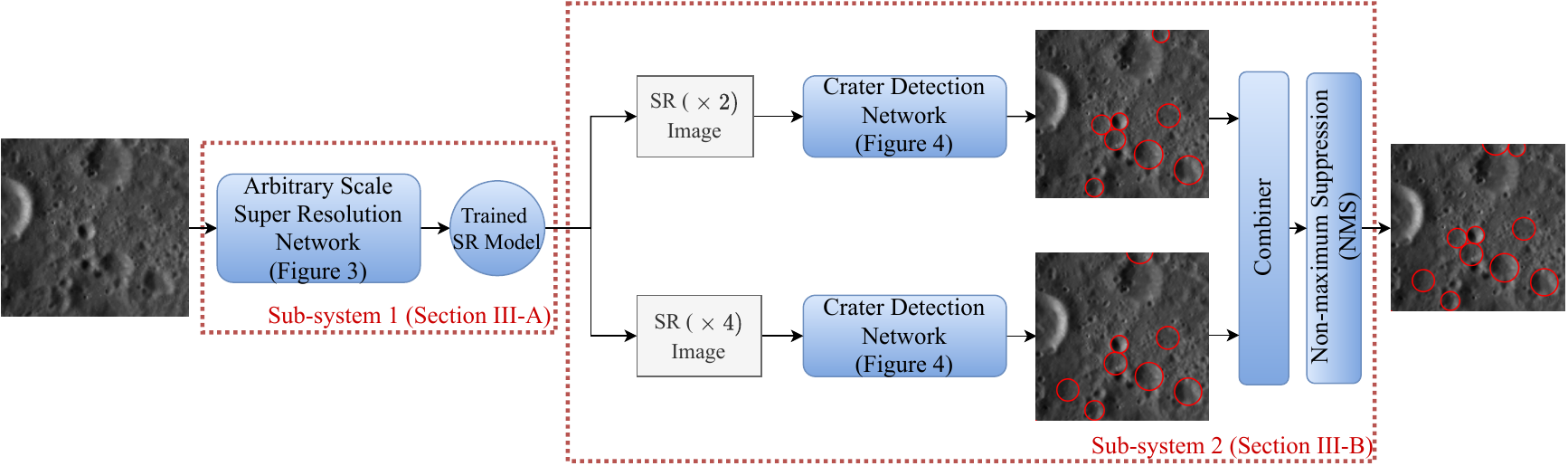}
    \caption{Overview of the SR-assisted crater detection system.}
    \label{fig:overview_sr_cda}
\end{figure*}

\section{Related Works}
\label{sec:related_works}

The existing works related to different aspects of this paper, super-resolution and crater detection, can be divided into three categories. In this section, the first two subsections briefly review the super-resolution systems for natural images and remote sensing images, and the last subsection reviews crater detection systems.

\subsection{Super-resolution for natural images}

The deep learning (DL) based super-resolution (SR) works can be typically divided into two categories: pre-upsampling-based and post-upsampling based approaches. 
The pre-upsampling-based approaches first upsample the LR images to the desired size and then restore high-frequency information using a deep neural network (DNN). 
Examples of pre-upsampling-based approaches are
SRCNN~\cite{dong2015image}, VDSR~\cite{Kim_2016_CVPR}, DRCN~\cite{Kim_2016_CVPR1}, and DRRN~\cite{Tai_2017_CVPR}. 
SRCNN proposed by Dong et al.~\cite{dong2015image} is the first convolutional neural network (CNN) based approach for SR. 
In this, interpolation is first done to get images of the desired size, and then three layers of CNN are used to restore the high-frequency information in these images. 
It outperformed the non-deep learning SR methods existing at that time. 
VDSR~\cite{Kim_2016_CVPR} proposed a very deep convolution neural network and introduced a residual connection to alleviate the training degradation problem in a deep network.
DRCN~\cite{Kim_2016_CVPR1} and DRRN~\cite{Tai_2017_CVPR} deployed recursive learning that helps to reduce the parameters while increasing the depth of the layers.
Though these pre-upsampling based approaches give better performance than the non-deep learning approaches, these are computationally expensive. 
Hence, most of the recent SR works follow post-upsampling approaches.  

In the post-upsampling approach, first features are learned from LR images, and then the learnable upsampling layer is used to construct the SR image. 
Examples of post-upsampling approaches are ESPCN~\cite{Shi_2016_CVPR}, EDSR~\cite{lim2017enhanced}, RDN~\cite{Zhang_2020_TPAMI}, and RCAN~\cite{zhang2018image}.
ESPCN proposed by Shi et al.~\cite{Shi_2016_CVPR} deployed learning-based upsampling, i.e., the sup-pixel layer, which, using convolution, generates multiple channels and then reshapes to perform upsampling.
Lim et al.~\cite{lim2017enhanced} proposed an enhanced deep residual network (EDSR), which modified the residual block by removing the batch normalization layers. 
This simple modification improved the performance substantially.
Zhang et al.~\cite{Zhang_2020_TPAMI} proposed a novel residual dense network (RDN) to efficiently extract the hierarchical features from low-resolution images. 
Specifically, they developed a residual dense block for utilizing abundant local features efficiently by dense connections.
Another work by Zhang et al. proposed a residual channel attention network (RCAN)~\cite{zhang2018image}, where residual in residual structure is used for efficient information flow and introduces a channel attention mechanism for improving representation ability of the CNNs.

Most existing works in SR are based on a single scale factor and only operate on integer scales. 
In applications such as remote sensing, multiple-resolution images are required to extract various levels of information from images~\cite{chen2023continuous}.
For example, for detecting craters of very diverse size ranges (diameters of few meters to more than thousand kilometers), we may need varying resolutions to meet the system's needed accuracy and space-time complexity. 
Training the SR network for different resolutions might be impractical in such applications as it takes huge training time and memory~\cite{lee2022local}. 
Hence it is necessary to design a single model which can super-resolve to any scale factor.
One of the earliest works to address arbitrary scale super-resolution was Meta-SR~\cite{hu2019meta}.
In Meta-SR, the Meta-upscale module was introduced that dynamically predicts the weights of filters using a fully connected network, and the predicted weights were used for upscaling to the desired resolution.
Inspired by implicit neural representation~\cite{sitzmann2020implicit}, Chen et al.~\cite{chen2021learning} proposed a local implicit image function (LIIF) for continuous magnification. 
LIIF utilized a multi-layer perceptron (MLP) as a local implicit function to predict the RGB value at arbitrary queried coordinates in HR image. The inputs to MLP are the features extracted from LR around the HR image coordinates, queried coordinates in the HR image, and cell size. 
For feature extraction, different fixed-scale super resolution methods such as RDN~\cite{Zhang_2020_TPAMI} and EDSR~\cite{lim2017enhanced} are used without their upscaling modules.
Another work, UltraSR~\cite{xu2021ultrasr} is another approach for arbitrary scale super-resolution which  blends spatial coordinates and periodic encoding to deal with spectral bias issues in MLP. 
To enhance the representational information of local implicit function, Lee et al.~\cite{lee2022local} introduced a local texture estimator (LTE) that characterized image texture in the Fourier domain.

\subsection{Super-resolution on remote sensing images}
Spatial resolution is critical for accurate inferences from remote sensing images. 
In past, typically high-resolution acquisition sensors were used for acquiring high-resolution images. 
However, high-resolution acquisition sensors are associated with constraints due to expensive hardware and transmission time. 
These limitations can be overcome by introducing super-resolution algorithms to synthesize high-resolution images from output of low-resolution sensors.
In recent years, significant progress has been made in remote sensing SR using deep learning techniques.
Lei et al.~\cite{lei2017super} introduced a local-global combined network (LGCNet) based on VDSR~\cite{Kim_2016_CVPR}, which learns multi-level features using a combination of local and global features to get the enhanced SR image.
The residual dense back projection network (RDBPN) introduced by Pan et al.~\cite{8732688} used residual, dense, up-projection, and down-projection modules to generate SR images with different scale factors.
Inspired by new concepts such as channel attention (CA), Gu et al.~\cite{rs11151817} introduced a residual squeeze and excitation block (RSEB) for utilizing different channel-inter-dependencies and giving enhanced SR image. 
A recursive-biLSTM approach was used in~\cite{rs11202333} to reduce the parameters and increase the receptive field, which effectively learns correlations of features and complementary information for effective image reconstruction.
A GAN-based SR is introduced in~\cite{rs11212578}, called dense residual generative adversarial network (DRGAN), which modified the Wasserstein GAN~\cite{arjovsky2017wasserstein} loss function, leading to better reconstruction and alleviating gradient vanishing problem. 
To reduce the effect of noise and artifacts, Jiang et al.~\cite{8677274} introduced an edge enhancement structure in conventional GAN.
Lu et al.~\cite{lu2019satellite} developed a multiscale residual neural network (MRNN) to investigate multiscale characteristics of objects in satellite imagery and restore high-frequency information of remote sensing images.
A frequency-domain scheme introduced in~\cite{ma2019achieving}, called wavelet transform combined with recursive ResNet (WTCRR), utilized various frequency bands through wavelet transform to reconstruct HR images.

\subsection{Crater detection}

Automatic crater detection algorithms (CDAs) can be divided into two categories: traditional and deep learning (DL) based algorithms. 
Traditional CDAs (e.g.~\cite{stepinski2009machine,yamamoto2015rotational}) typically first
identify potential crater features, such as edges and contours, and then use these features for crater detection. 
In the past few years, there has been a shift towards employing deep learning-based systems for crater detection. 
Deep learning-based systems learn the features on their own and do not need to explicitly extract crater features. 
One of the pioneering works of DL-based CDAs was by Silburt et al.~\cite{silburt2019lunar}. 
In this, authors first utilized the U-Net architecture~\cite{ronneberger2015u} to segment crater and non-crater regions from the input digital elevation map (DEM) data. 
Then, the segmented image was passed to the template-matching algorithm for extracting craters' locations and radii.
Similarly, some other works (e.g.,~\cite{delatte2019segmentation,lee2019automated,lee2021automated,wang2020effective,mao2022coupling}) also utilized U-Net variants for the crater detection task.
Another set of studies have used an object detection framework~\cite{TEWARI2022105500,ali2020automated,tewari2023automatic,yang2020lunar,yang2022progressive} for crater detection. 
For example, Ali-Dib et al.~\cite{ali2020automated} utilized Mask R-CNN framework~\cite{he2017mask} for crater detection.
Detailed surveys of deep learning-based crater detection works are presented in DeLatte et al.~\cite{delatte2019automated} and Tewari et al.~\cite{tewari2023deep}. 

To the best of our knowledge, only a single work utilizes a super-resolution algorithm for crater detection task (Grassa et al.~\cite{la2023yololens}).
It consists of two sub-networks: Generator and YOLOv5~\cite{redmon2016you}. 
A Generator is a super-resolution framework that is used to predict high-resolution images from low-resolution counterparts. 
The super-resolution output is fed to the object detection framework (YOLOv5) for detecting craters.
Grassa et al.~\cite{la2023yololens} used a refinement learned network (RLNet)~\cite{la2022adversarial} for super-resolution.
In contrast, the system proposed in this paper is based on an arbitrary scale super-resolution method that eliminates the need to train multiple models for different scales. 
Further, it utilizes a two-stage object detection framework (Mask R-CNN~\cite{he2017mask}) for crater detection. Therefore, the proposed system is complementary to the system presented in~\cite{la2023yololens} and will help to understand the impact of arbitrary scale super-resolution on crater detection performance.

\section{Methodology}
\label{sec:methods}

The proposed system consists of two subsystems. 
The first subsystem uses an arbitrary scale super-resolution approach for improving the resolution of lunar optical images (Section~\ref{subsec:Enhancing the resolution of the optical images}). 
The second subsystem utilizes the predicted SR images of multiple resolution in training the crater detection networks. 
Further, detected craters from the crater detection model are processed to remove duplicate and partial craters and convert craters' location information from pixel coordinates to geographic coordinates (Section~\ref{subsec:Crater Detection on SR predicted images}).
Figure~\ref{fig:overview_sr_cda} shows an overview of the proposed SR-assisted crater detection system.

\subsection{Enhancing lunar optical images' resolution}
\label{subsec:Enhancing the resolution of the optical images}

\begin{figure*}[ht!]
    \centering
    \includegraphics[width=\textwidth]{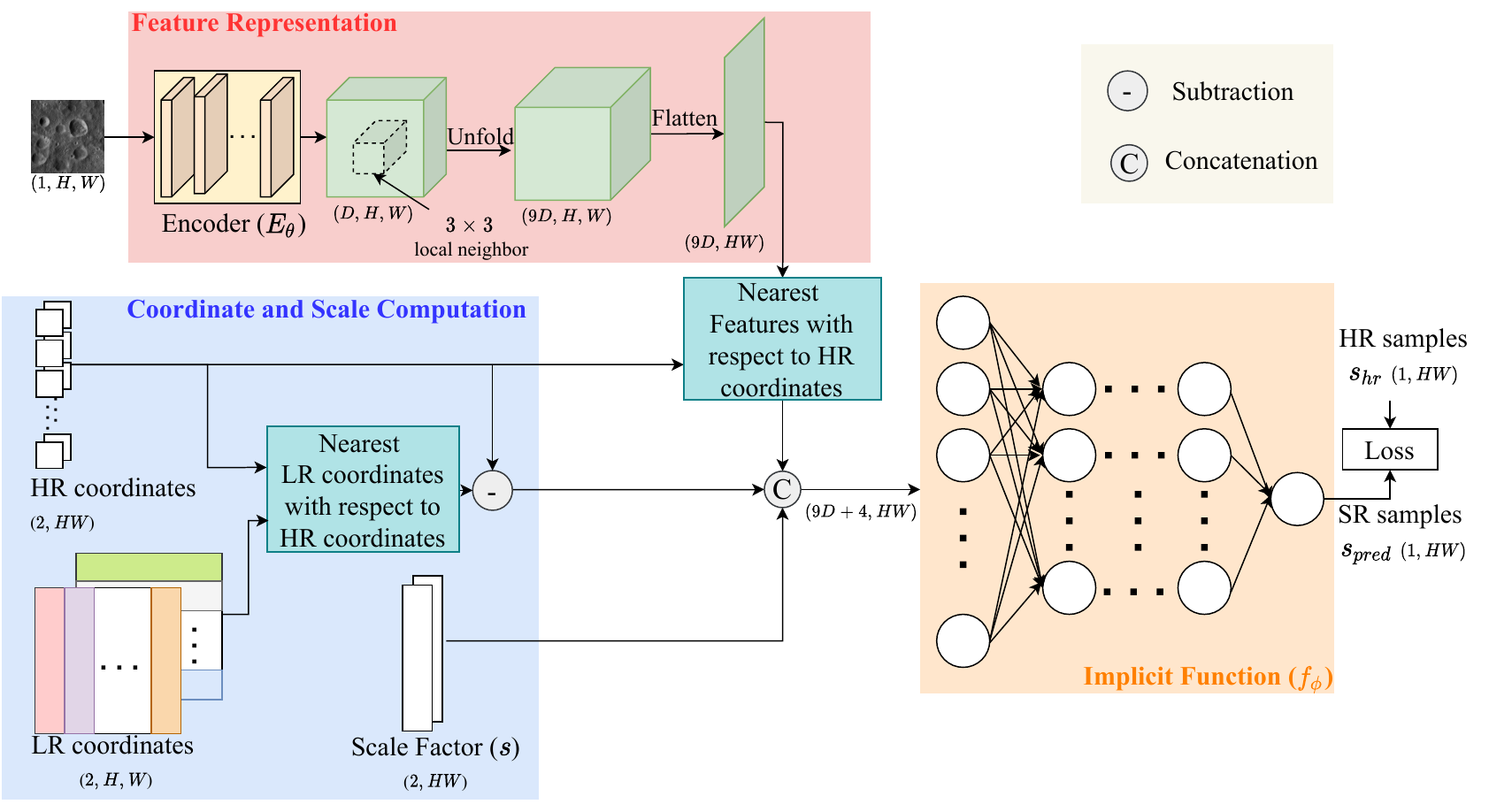}
    \caption{Overview of arbitrary scale super-resolution approach based on local implicit image function (LIIF)~\cite{chen2021learning}.}
    \label{fig:liif_overview}
\end{figure*}

The proposed system improves the resolution of lunar optical images with continuous magnification by using an arbitrary scale super-resolution approach, local implicit image function (LIIF)~\cite{chen2021learning}. 
Figure~\ref{fig:liif_overview} and Figure~\ref{fig:overview_sr_cda} (Sub-system 1) show an overview of this method. 

LIIF mainly consists of feature representation, coordinate and scale computation, and implicit function ($f_\phi$).
Feature representation encompasses an encoder and feature unfolding. Encoder, $E_\theta$ (with $\theta$ as its parameters), is used to learn the deep representation from its (?) low-resolution counterpart. It comprises a  convolutional neural network (CNN) and transformer modules. 
$E_\theta$ is used to extract the feature representation denoted as $M\in \mathds{R}^{\{D\times H\times W\}}$, from the input $I_{LR}\in \mathds{R}^{\{C\times H\times W\}}$. 
Here, $H$ and $W$ are the image's width and height, $D$ is the number of features, and $C$ is the number of channels in the input. 
Chen et al. used RDN~\cite{Zhang_2020_TPAMI} and EDSR-baseline~\cite{lim2017enhanced} without their upscale modules for the encoder. 
In our work, apart from RDN and EDSR-baseline, SwinIR~\cite{liang2021swinir} is also used for encoder. 
RDN and EDSR are based on convolution neural networks, whereas SwinIR is mainly based on a swin transformer~\cite{liu2021swin}. 

Feature information is further enhanced by performing feature unfolding on the feature map, $M$, and obtaining $\tilde{M}$. $\tilde{M}$ is formed by concatenating the $3\times3$ neighboring feature maps of $M$.

\begin{align}
    \tilde{M}_{i,j} = Concat(\{M_{i+m,j+n}\}_{m,n \in\{-1,0,1\}})
\end{align}

The output of feature unfolding is then flattened, and the nearest features concerning high-resolution (HR) coordinates are estimated. 

The 2D coordinates in the continuous image domain are used for continuous representation, as shown in the coordinate and scale computation block in Figure~\ref{fig:liif_overview}. 
First, the nearest coordinates of the LR image with respect to HR coordinates are calculated. 
The difference between HR and the nearest LR coordinates is computed to estimate the relative locations. 
These relative coordinates are subsequently fed to the implicit function ($f_\phi$) to calculate SR pixels. In addition to relative coordinates, scale factor information is incorporated. 
The scale factor information can be useful for achieving a continuous representation in a desired resolution.

The nearest feature representation, relative coordinates, and scale factor information are concatenated and passed through the implicit function ($f_\phi$). 
The implicit function is implemented using a multi-layer perceptron (MLP) with $5$ layers that computes the SR pixel corresponding to a particular HR coordinate. 
This MLP has ReLU~\cite{glorot2011deep} activation after each layer and each layer's dimension is set to 256. 
The HR pixel value for the image at a particular HR coordinate, $x_q$, is given by

\begin{align}
    I(x_q)&=f_{\phi}(z^*,x_q-p(z^*))
\end{align}
where $z^*$ is the nearest latent code with respect to the HR coordinates in the feature map ($\tilde{M}$) and $p(z^*)$ is the 2D coordinate of $z^*$.

A discontinuous prediction occurs when $x_q$ moves and $z^*$ suddenly changes; two close coordinates get predicted by different latent codes. 
To address this, a local ensemble strategy is used in the $2\times2$ latent code neighborhood. 
Each prediction is weighted by the area of the rectangle between the query coordinate and $p(z^*)$. 
The weight will be higher when the $p(z^*)$ is closer to the query coordinate.

\begin{align}
    I(x_q)&= \sum_{t\in\{00,01,10,11\}} \frac{S_t}{S}. f_{\phi}(z_t^*,x_q-p(z_t^*)),
\end{align}
where $z_t^*$ is the nearest latent code. 
$z_{00}^*,z_{01}^*,z_{10}^*,$ and $z_{11}^*$ correspond to the latent codes at the top left, top right, bottom right, and bottom left positions, respectively. 
$S_t$ is the area of rectangle between $x_q$ and $p(z_{t_1}^*)$ coordinates, where $t_1$ corresponds to diagonal position with respect to $t$. $S =\sum_{t} S_t$, is used for weight normalization.

Finally, $L_1$-loss is calculated between predicted intensity values ($s_{pred}$) and ground truth intensity values ($s_{hr}$) (Section~\ref{subsec:Dataset preparation}) for optimizing weights of the neural network.

\subsection{Crater detection on super-resolved images}
\label{subsec:Crater Detection on SR predicted images}

\begin{figure*}[ht!]
    \centering
    \includegraphics[width=\textwidth]{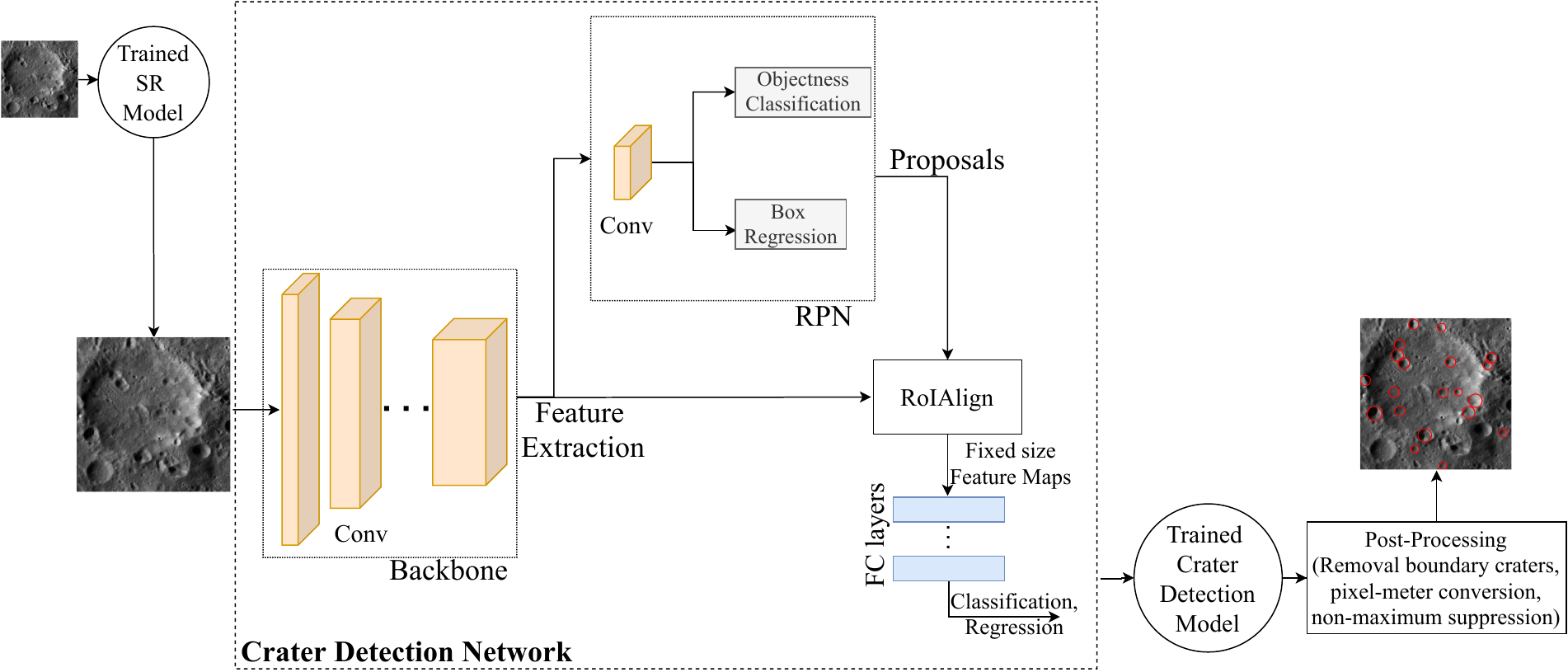}
    \caption{Crater detection network from predicted images from SR model.}
    \label{fig:cda_sr_arb}
\end{figure*}

The predicted images from SR model are used for training and testing the crater detection network.
The overview of the crater detection network from predicted super resolved images of a single scale factor (e.g., SR ($\times 2$)) is shown in Figure~\ref{fig:cda_sr_arb} and Figure~\ref{fig:overview_sr_cda} (part of Sub-system 2).
For crater detection network, we have utilized the Mask R-CNN framework~\cite{he2017mask}. 
Mask R-CNN mainly consists of feature extraction network, region proposal network (RPN), and detection network. 
First, features from the image are extracted using a feature extraction network based on convolutional neural network modules. 
The extracted features are passed to the RPN and ROIAlign. 
RPN consists of convolutional and fully connected (FC) layers and calculates locations of potential craters. 
The location information from RPN with extracted features from the feature extraction network is provided to the ROIAlign. ROIAlign calculates the fixed-size feature map of each potential crater. 
This fixed-size feature map is further passed to the detection network for removing non-crater regions and refining crater locations. 
Detection network consists of FC layers. 
It should be noted that in our work, we have used Mask R-CNN for crater detection and not for segmentation.

Similar to Tewari et al.~\cite{TEWARI2022105500}, the post-processing steps include removing the boundary craters, non-maximum suppression (NMS), and pixel-to-meter conversion. 
The complete optical mosaic of the lunar surface is divided into overlapping patches (sub-images).
Hence, craters around the boundary are split into multiple parts. Consequently, a single patch may contain partial craters around the boundary. Detection of these craters is undesirable; hence, a boundary crater removal step is needed to eliminate the partial craters. 
Pixel-to-meter conversion converts the pixel coordinate information to geo-graphic coordinate information for obtaining the global location. 
Finally, non-maximum suppression is applied to remove the duplicate craters.
In NMS, the highest confidence score crater is selected, and then intersection over union (IoU) with other craters is calculated. Craters with an IoU greater than the IoU threshold are removed. This process is repeated till either all craters are selected or suppressed.

\section{Experimental Results and Discussions}
\label{sec:experimental results}

The SR sub-system is implemented in PyTorch~\cite{paszke2019pytorch} library on RTX-$5000$ GPU. 
The nomenclature for different models is <encoder-name>-LIIF ($\times$scale factor). 
For example, SwinIR~\cite{liang2021swinir} encoder with scale factor $4$ is denoted as SwinIR-LIIF ($\times$4).
Total epochs are 300 with Adam optimizer~\cite{kingma2014adam}, setting $\beta_1$=$0.9$, $\beta_2$=$0.999$, and $\epsilon$=$10^{-8}$. 
Batch size is $8$ and learning rate is $10^{-4}$, with decay by half after $200$ epochs. The network is trained using the $L_1$ loss function. 

The predicted images from different SR models are used in crater detection. In crater detection, we have used Mask R-CNN implementation by Matterport~\cite{matterport_maskrcnn_2017}, which used Keras with the TensorFlow backend.
To extract the feature in Mask R-CNN framework, in the backbone, we have used ResNet-50~\cite{he2016deep} architecture with feature pyramid network (FPN)~\cite{lin2017feature}. 
The following augmentations are processed to enhance the variation of crater features in training.
The augmentation includes rotation of images by $90^\circ$, $180^\circ$, and $270^\circ$, change of intensity values of optical images by $80\%$ to $150\%$ of original values, Gaussian blur with sigma value $0$ to $5$, and both horizontal and vertical flipping from the center of the images. 
In our experiments on crater detection sub-system, stochastic gradient descent (SGD) optimizer with a learning rate of 0.01 is utilized and total $40$ epochs are used for training. 

\subsection{Dataset preparation}
\label{subsec:Dataset preparation}

For the SR task, we have used digital optical images collected by the Kaguya terrain camera~\cite{haruyama2008global}.
It has a resolution of $\sim$$7.4$ m/pixel ($4096$ pixel/degree) and a bit depth of $16$ bits. 
In our study, we downsampled optical images to a resolution of $50$ m/pixel using bicubic interpolation in ArcMap. 
As a result, optical images of $50$ m/pixel served as high-resolution (HR) ground truth data, having a size of $1816\times1816$ pixels.
The study area swathed the entire longitude range from -$180^{\circ}$ to $180^{\circ}$ and a latitude range of $\pm 60^{\circ}$. 
A total of $4800$ optical images from the Kaguya data archive~\cite{haruyama2008global} are collected. 
We excluded images that have missing pixel values. 
Further, we observed that $96.99\%$ of the optical images had a dynamic range of $\leq 3000$ meters. 
Hence, our work focuses on optical images with a dynamic range of $\leq 3000$ meters.
The following selection process is used to cover the latitude range of $\pm 60^{\circ}$ and the longitude range of  -$180^{\circ}$ to $180^{\circ}$: 
\begin{enumerate}
    \item For every $3^{\circ}$ interval across the complete longitude range, three optical images are collected.
    \item The latitude range is divided into three intervals: [-$60^{\circ}$, -$21^{\circ}$], [-$21^{\circ}$, $21^{\circ}$], and [$21^{\circ}$, $60^{\circ}$].
    \item The lunar surface's north, mid, and south regions are covered by choosing the optical images for each latitude interval following above two steps. 
    \item Thus, we acquired 216, 24, and 120 optical images for SR training, validation, and testing. 
\end{enumerate}

The training setup is similar to the existing works~\cite{lim2017enhanced,Zhang_2020_TPAMI}. Training images have a size of $(1, H_{tr}, W_{tr})$, where $H_{tr}$ and $W_{tr})$ represent height and width of the training images. 
In our case, it is $(1, 1816, 1816)$.  
We randomly cropped patch size $48s \times 48s$ from the high-resolution (HR) image. 
The upscaling factor `s' is sampled from the uniform distribution $s \sim \mathcal{U}(1,4)$. The cropped HR patches are downsampled using bicubic interpolation. For bicubic interpolation, we have utilized the resize function in Torchvision~\cite{paszke2019pytorch} and then augmented randomly by horizontal flip, vertical flip, $90^{\circ}$ rotation, brightness variation, and contrast variation. For ground truth, $48^2$ pixel samples ($s_{hr}$) and corresponding coordinates are used from HR ($x_{hr}$).

For crater detection task, we utilized the lunar optical images generated by Tewari et al.~\cite{TEWARI2022105500}. 
They utilized the optical images captured from the lunar reconnaissance orbiter (LRO) wide-angle camera (WAC) on the LRO~\cite{robinson2010lunar}, which has a resolution of $100$ m/pixel.
The study region considers longitude from -$180^\circ$ to $180^\circ$ and latitude from -$60^\circ$ to $60^\circ$. The training region spans longitude from -$180^\circ$ to $60^\circ$ and latitude from -$60^\circ$ to $60^\circ$.
The complete mosaic is divided into overlapping patches of size $1024\times1024$ with $50\%$ overlap. 
To obtain the low-resolution (LR) images, we downsample the original image with a scale factor of $4$, resulting in LR images having a size of $256\times256$ and an effective resolution of $400$ m/pixel. 
Now using arbitrary scale SR models, we predicted the SR counterpart of LR images. 
We predicted the SR counterpart with scale factor $2$ (SR ($\times2$)) and scale factor $4$ (SR ($\times4$)). SR ($\times2$) has a size of $512\times512$ with an effective resolution of $200$ m/pixel;  SR ($\times4$) has a size of $1024\times1024$ with an effective resolution of $100$ m/pixel. 

In our work, we have used craters present in Robbins et al.~\cite{robbins2019new} catalog of diameter size $5$-$10$ km for training and evaluation of crater detection work.
The total number of images for training and testing the crater detection network are $8552$ and $5041$, with the number of craters being $27319$ and $21761$, respectively. 
Further, the training set is divided into train and validation; the train set contains $6933$, and the validation set contains $1619$ images.

\subsection{Evaluation metrics}
\label{subsec:evaluation metrics}

Precision, recall, and F$_1$-Score are used to evaluate the crater detection performance. 
Precision (P) tells how many detected craters match with ground truth, with respect to the total number of craters detected by the model. 
Recall (R) tells how many detected craters match with the ground truth, with respect to total ground truth craters.
$F_1$-Score balances precision and recall, and it is the harmonic mean of precision and recall. 
These are defined as (\cite{baeza1999modern}): 
\begin{align}
\text{P} &= \frac{T_P}{T_P + F_P}\times100\\
\text{R} &= \frac{T_P}{T_P + F_N}\times100\\
\text{F$_1$-score} &= \frac{2 \times \text{P} \times \text{R}}{\text{P} + \text{R}},
\end{align}
where $T_P$ denotes true positives, i.e., the number of craters detected by a model that match with the ground truth, $F_P$ denotes false positives, i.e., the number of craters detected by the model that are not marked in the ground truth, and  $F_N$ defines the number of craters that are present in the ground truth but are not detected by the model.

\subsection{Optimization of post-processing parameters}
\label{subsec:Determine the optimal post-processing parameters}

This section explains the optimization of post-processing parameters in our work. 
The post-processing parameters include the parameter controlling removal of craters around the boundary ($m$), NMS confidence score ($s$), and NMS IoU threshold ($\tau$). 
The parameters $m$, $s$, and $\tau$ are simultaneously optimized with values of $m \in \{0,5,10,15\}$, $s \in \{0,0.6,0.7,0.8,0.9\}$, and $\tau \in \{0.1,0.2,0.3,0.4,0.5,0.6\}$. 
This yields a total of $4\times5\times6$ cases. 
The optimal parameters were chosen based on the best $F_1$-score for the validation set and utilized in evaluating the performance for test set.
For easy understanding, Table~\ref{tab:Rmv_bdry_sr}, Table~\ref{tab:robbins_NMS_cf_sr}, and Table~\ref{tab:robbins_NMS_sr} show the variation of one parameter, while fixing others. 
This section presents the parameter optimization for the SwinIR-LIIF ($\times4$) model. 
A similar analysis was done for other models.

Table~\ref{tab:Rmv_bdry_sr} shows the effect of removing the boundary craters. This step is needed to remove the partial craters around the boundary of the images. Increasing values of $m$ removed more partial craters, reducing the number of false positives, hence, improving the precision. 
However, it also removes some of the true positives, which causes decrease in recall value. We get the optimal value in the validation set at m equal to $5$, where we get the best $F_1$-score ($63.02\%$).

\begin{table}[ht!]
	\centering
	\caption{Impact of Removing Boundary Craters ($\tau = 0.5$, $s = 0.7$). }\label{tab:Rmv_bdry_sr}
	\begin{tabular}{p{0.03\textwidth} p{0.11\textwidth} p{0.09\textwidth}p{0.11\textwidth}} 
		\hline
		$m$ & Precision (\%) & Recall (\%) & F$_1$-score (\%) \\ [0.5ex] 
		\hline
	    0 & 66.90 & 57.57 & 61.89 \\ 
		1 & 69.58 & 57.10 & 62.72 \\ 
		5 & 71.21 & 56.51 & 63.02 \\
		10 & 71.41 & 56.23 & 62.92 \\
		15 & 71.50 & 55.95 & 62.78 \\
		\hline
	\end{tabular}
\end{table}

Table~\ref{tab:robbins_NMS_cf_sr} shows the effect of confidence score threshold ($s$) in crater detection. 
A crater with high confidence score is more certain to be a true crater.
Increasing the confidence threshold will eliminate the less confident craters but may also remove some true craters; hence, the precision will increase, and recall will reduce. The best $F_1$-score we get is at $s$ = $0.7$.

\begin{table}[ht!]
	\centering
	\caption{Impact of using confidence score threshold ($m = 5$, $\tau = 0.5$).}\label{tab:robbins_NMS_cf_sr}
	\begin{tabular}{p{0.03\textwidth} p{0.11\textwidth} p{0.09\textwidth}p{0.11\textwidth}} 
		\hline
		$s$ & Precision (\%) & Recall (\%) & F$_1$-score (\%) \\ [0.5ex] 
		\hline
		0 & 58.27 & 64.68 & 61.31 \\ 
		0.6 & 64.42 & 60.55 & 62.43 \\
		0.7 & 71.21 & 56.51 & 63.02 \\
		0.8 & 78.51 & 50.91 & 61.77 \\
		0.9 & 87.52  & 41.63 & 56.42 \\
		\hline
	\end{tabular}
\end{table}

The NMS IoU threshold ($\tau$) is utilized to eliminate duplicate craters (Table~\ref{tab:robbins_NMS_sr}). The significance of the NMS IoU threshold ($\tau$) becomes apparent by comparing the last row (without NMS) with other rows; it is evident that the precision is drastically decreasing without NMS. 
Reducing $\tau$ decreases the number of false positive craters, resulting in improved precision. However, we found a drop in precision at $\tau$ =0.2 due to the simultaneous reduction in true positive craters. 
While decreasing the $\tau$ value removes more duplicate craters, it also removes some true craters. Consequently, a decrease in true positive craters and an increase in false negative craters leads to a reduced recall value. 
In most scenarios, reducing the $\tau$ value increases the precision and decreases the recall. We found the optimum $F_1$-score of 63.02\% at $\tau$ =0.5.

\begin{table}[ht!]
	\centering
	\caption{Impact of using NMS IoU threshold ($m = 5$, $s = 0.7$).}\label{tab:robbins_NMS_sr}
	\begin{tabular}{p{0.1\textwidth} p{0.10\textwidth} p{0.09\textwidth}p{0.1\textwidth}} 
		\hline
		$\tau$ & Precision (\%) & Recall (\%) & F$_1$-score (\%) \\ [0.5ex] 
		\hline
		0.1 & 71.26 & 55.13 & 62.17 \\ 
		0.2 & 71.27 & 56.12 & 62.79 \\
		0.3 & 71.29 & 56.36 & 62.95 \\
		0.4 & 71.29 & 56.45 & 63.01 \\
		0.5 & 71.21  & 56.51 & 63.02 \\
		0.6 & 71.07  & 56.55 & 62.98 \\  
		Without NMS & 19.67 & 65.59 & 30.26 \\
		\hline
	\end{tabular}
\end{table}

In a similar manner we have calculated the optimum values of $m$, $\tau$, and $s$ for other models (Table~\ref{tab:robbins_m_tau_s_sr_all}).

\begin{table}[ht!]
    \centering
    \caption{Optimal values of post processing parameters ($m, \tau, s$) for different approach.}
    \label{tab:robbins_m_tau_s_sr_all}        
    \begin{tabular}{p{0.07\textwidth}p{0.05\textwidth}p{0.07\textwidth}p{0.07\textwidth}p{0.07\textwidth}} \hline
    Parameters & LR & RDN-LIIF & EDSR-LIIF & SwinIR-LIIF  \\  \hline
    $m$ & 0 & 5 & 5 & 5 \\ 
    $\tau$ & 0.4 & 0.5 & 0.5 & 0.5 \\
    $s$ & 0.8 & 0.7 & 0.6 & 0.7 \\ \hline
    \end{tabular}
\end{table}

\subsection{Crater detection evaluation}
\label{subsec:Crater Detection Evaluation}

To evaluate the performance of the crater detection, we have used Robbins et al.~\cite{robbins2019new} catalog for crater diameter range $5-10$ km. Table~\ref{tab:SR_methods_encoders} compares the crater detection model trained on LR images with the crater detection model trained on the images predicted by SR models. 
It can be observed that all SR models (EDSR-LIIF, RDN-LIIF, and SwinIR-LIIF) are outperforming LR by a significant margin. 
The SwinIR-LIIF model has the best $F_1$-score among all SR models. Consequently, for further analysis, we have used the SwinIR model. 
Compared to the LR model, the SwinIR-LIIF model has a precision, recall, and $F_1$-score improvement of $3.44\%$, $6.71\%$, and $5.65\%$, respectively.
Therefore, we concluded that integrating SR approaches in crater detection work enhances the overall performance of crater detection.

\begin{table}[ht!]
    \centering
    \caption{Performance of crater detection using SR models at scale factor 4 (the best result is marked as \textbf{bold}).}\label{tab:SR_methods_encoders}
 {
    \begin{tabular}{p{0.1\textwidth}p{0.1\textwidth} p{0.08\textwidth}p{0.1\textwidth}} 
        \hline
        Models & Precision (\%) & Recall (\%) & F$_1$-score (\%) \\ [0.5ex]
        \hline
        LR & 70.83 &	51.16 &	59.41 \\ 
        \hline
        EDSR-LIIF & 66.47 &	\textbf{59.55} &	62.82  \\ 
        RDN-LIIF  & 72.01 &	56.31 &	63.20  \\
        SwinIR-LIIF  & \textbf{74.27} & 57.87 & \textbf{65.06} \\  
        \hline
    \end{tabular}
 }
\end{table}

\begin{figure*}[ht!]
    \centering
    \includegraphics[width=0.8\textwidth]{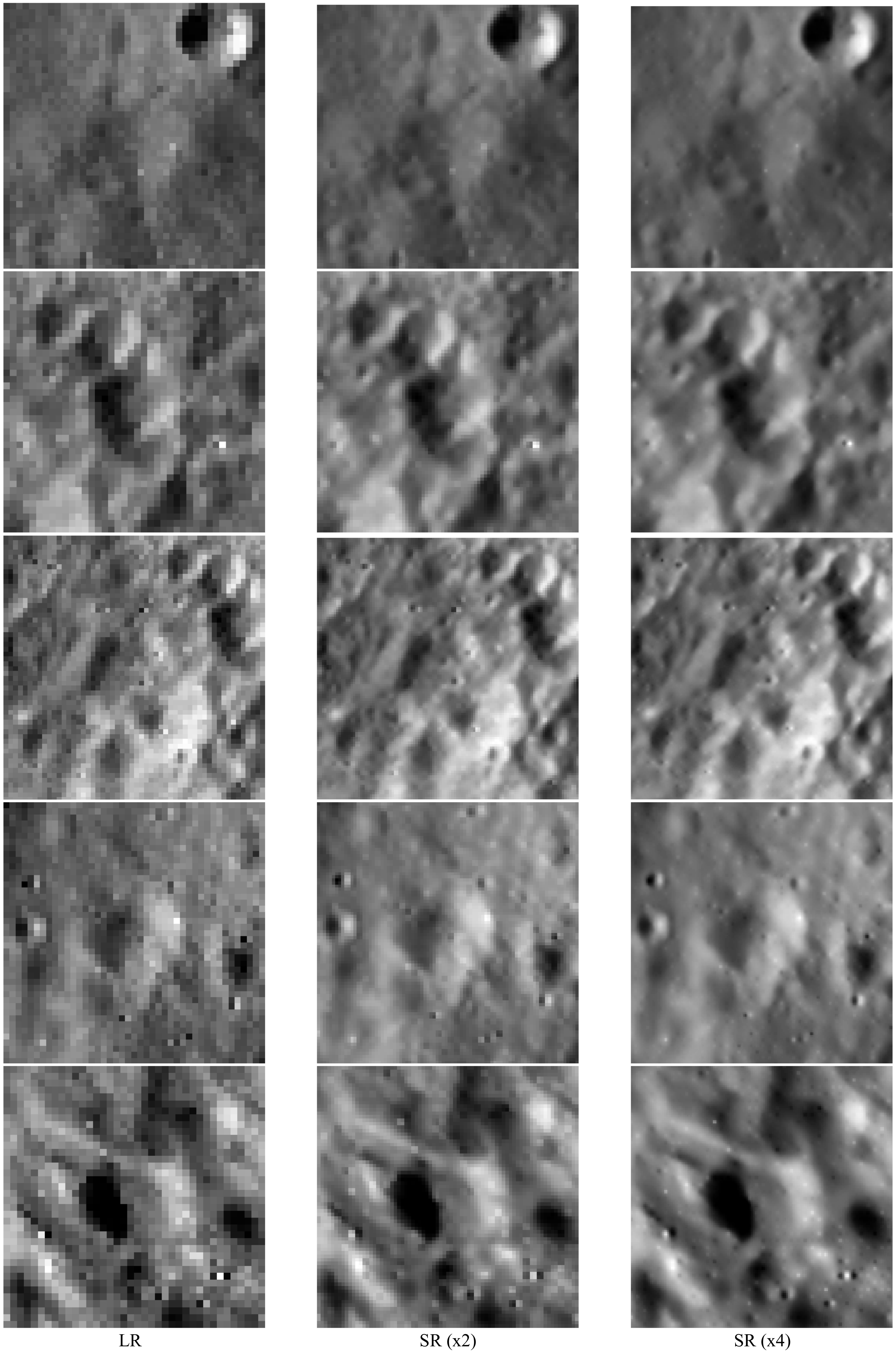}
    \caption{Visual inspection of the sample images. LR is the low resolution image, SR ($\times$2) is the prediction on SR images with scale factor 2, SR ($\times$4) is the prediction on SR images with scale factor 4.}
    \label{fig:visual_craters_sr}
\end{figure*}

Figure~\ref{fig:visual_craters_sr} displays visualization of sample images for the SwinIR-LIIF model with scale factors $2$ ($\times2$) and $4$ ($\times4$). 
It is evident that the SR process can recover finer details, leading to improved performance in crater detection, specifically in densely populated crater regions and degraded craters' scenarios. 
For example, in the LR image shown in the last row of Figure~\ref{fig:visual_craters_sr}, crater and non-crater regions are challenging to distinguish, whereas, in the corresponding SR images, the differentiation between the two regions is significantly more apparent.

\begin{figure*}[ht!]
    \centering
    \includegraphics[width=0.9\textwidth,trim= 0 0.25cm 0 0,clip]{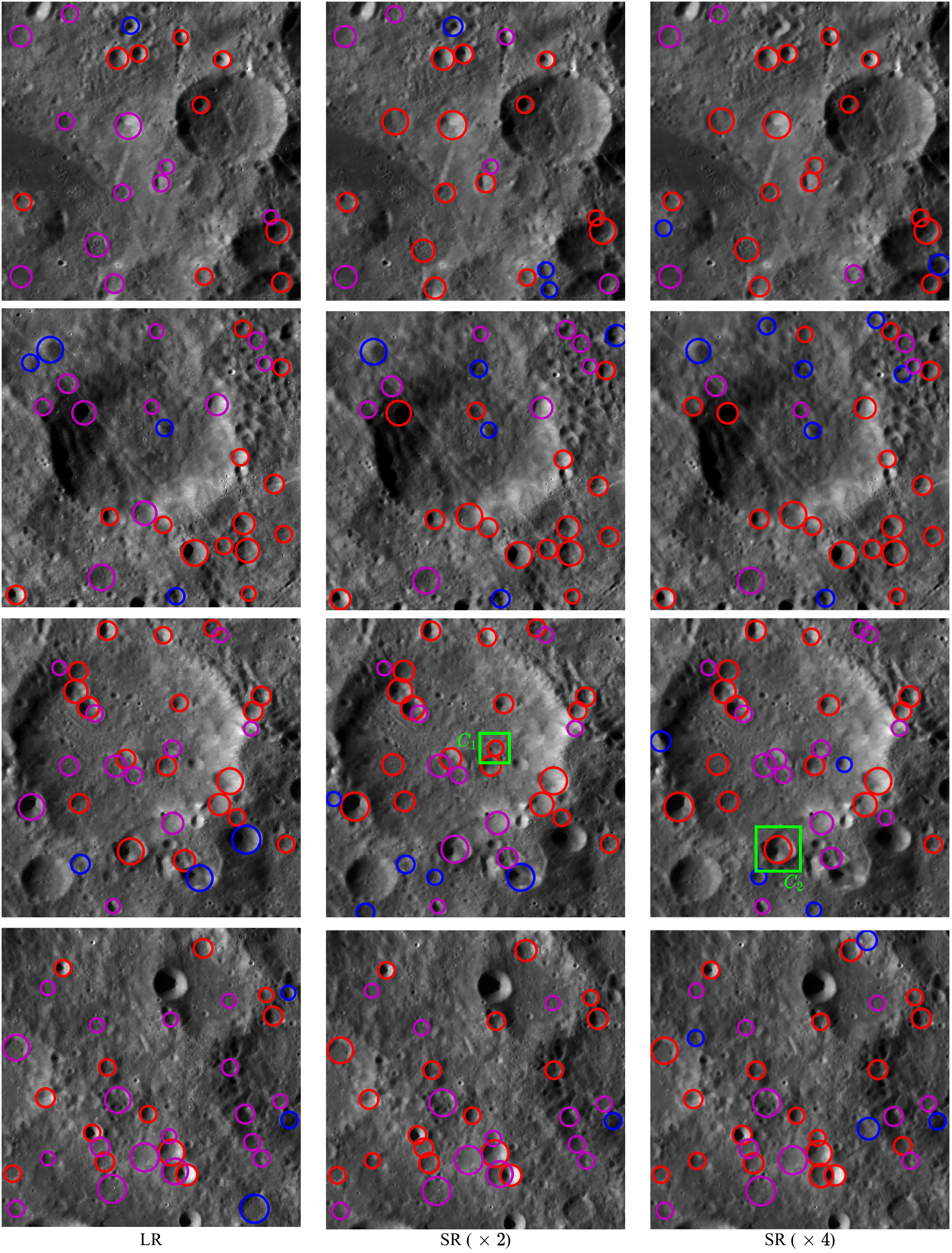}
    \caption{Visualization of the craters detected from LR, SR ($\times2$), and SR ($\times4$) images (from left to right). Color coding: Red: True positive, Blue: False positive, and Purple: False negative.}\label{fig:sample_cda_sr}
\end{figure*}

Figure~\ref{fig:sample_cda_sr} shows visualization of the craters detected from LR, SR ($\times2$), and SR ($\times4$) images. 
Color coding for the craters is as follows: red denotes true positive craters, blue denotes false positive craters, and purple denotes false negative craters. 
It can be observed that SR models give more true positive craters and fewer false negatives compared to their LR counterparts. 
In complex scenarios (such as overlapping crater regions) also the predicted images from the SR models have better crater detection performance than the LR model (for example the last row in Figure~\ref{fig:sample_cda_sr}). 
In some cases, few of the craters detected in the SR images are marked as a false positive but by visual inspection they seem to be true craters (for example, in the first row, second column of Figure~\ref{fig:sample_cda_sr}, there is are three craters marked in blue and all three of them appear to be actual craters based on visual inspection). 
Such cases are likely to occur for craters whose cataloged diameters are slightly lesser than 5 km because for the performance evaluation we have restricted crater diameter range to 5-10 km in the catalog. 

\subsubsection{Combination of results from multiple resolutions}
\label{subsubsec:Impact of combining crater detection outputs from multiple models}

Our observations indicate that in some scenarios, craters are detected by one SR model (e.g., SR ($\times2$) but undetected by another model (e.g., SR ($\times4$)) and vice versa. 
For example, in the fourth row of Figure~\ref{fig:sample_cda_sr}, the crater denoted as $C_1$
is detected in a predicted image from the SR ($\times2$) model but undetected in SR ($\times4$). Similarly, a crater denoted as $C_2$ is detected in a predicted image from the SR ($\times4$) model but undetected in SR ($\times2$).
This phenomenon may be due to different resolutions containing distinct representations in deep learning models; consequently, in some scenarios, one model can detect craters while another fails to detect them. It highlights the possible advantages of using multiple resolutions to improve the detection of craters. 
Therefore, we combine the outputs of crater detection networks trained on multiple resolutions (e.g., SR ($\times2$) and SR ($\times4$)). 
We have investigated various combinations of models to analyze their impact on crater detection performance, as shown in Table~\ref{tab:SR_combine}. 
For example, in the fourth row of Table~\ref{tab:SR_combine}, the entry LR, SR ($\times2$) denotes the combined detection results of the LR and SR ($\times2$) models.
Combining the detection results of multiple models leads to increased detection, consequently increasing the recall. It can be observed that in any combination of models, the recall is higher than in a single model. 
The highest recall is obtained by combining the results of three models: LR, SR ($\times2$), and SR ($\times4$) (Table~\ref{tab:SR_combine}, last row). 

However, combined results also increase the number of false positives, consequently reducing the precision value. Among all combination results, we get the best precision on the combination of SR ($\times2$) and SR ($\times4$) models. 
Also, combined results of SR ($\times2$) and SR ($\times4$) have the best F$_1$-score with a 5.76\% improvement compared to the LR model.
Hence, combining results from SR models will be the optimal approach.
Additionally, combining results from multiple models will be more beneficial in applications prioritizing recall, such as hazard avoidance.

\begin{table}[ht!]
    \centering
    \caption{Performance improvement by combining results from multiple models. The best value for combined results is highlighted in \textbf{bold}.}\label{tab:SR_combine}
 {
    \begin{tabular}{p{0.16\textwidth} p{0.08\textwidth} p{0.05\textwidth}p{0.08\textwidth}} 
        \hline
        Models & Precision (\%) & Recall (\%) & F$_1$-score (\%) \\ [0.5ex]
        \hline
        LR & 70.83 &	51.16 &	59.41 \\ 
        SR ($\times2$) & 70.58 & 59.68 & 64.67  \\ 
        SR ($\times4$) & 74.27 &	57.87 &	65.06  \\ \hline
        LR, SR ($\times2$) & 63.56 & 63.24 & 63.40      \\
        LR, SR ($\times4$) & 65.40 &	62.70 &	64.02\\   
        SR ($\times2$), SR($\times4$) & \textbf{65.73} &	64.63 &	\textbf{65.17} \\   
        LR, SR ($\times2$), SR ($\times4$) & 60.64 & \textbf{66.85} & 63.59  \\ \hline
    \end{tabular}
 }
\end{table}

\subsubsection{Localization performance}

Crater localization performance refers to how accurately a crater detection system can align the detected craters' location with ground truth craters. It is calculated using intersection over union (IoU) (in percentage), quantifying the overlap between a detected true crater and a ground truth crater. A higher IoU value infers a better localization performance.

\begin{figure}[ht!]
    \centering    \includegraphics[width=0.442\textwidth]{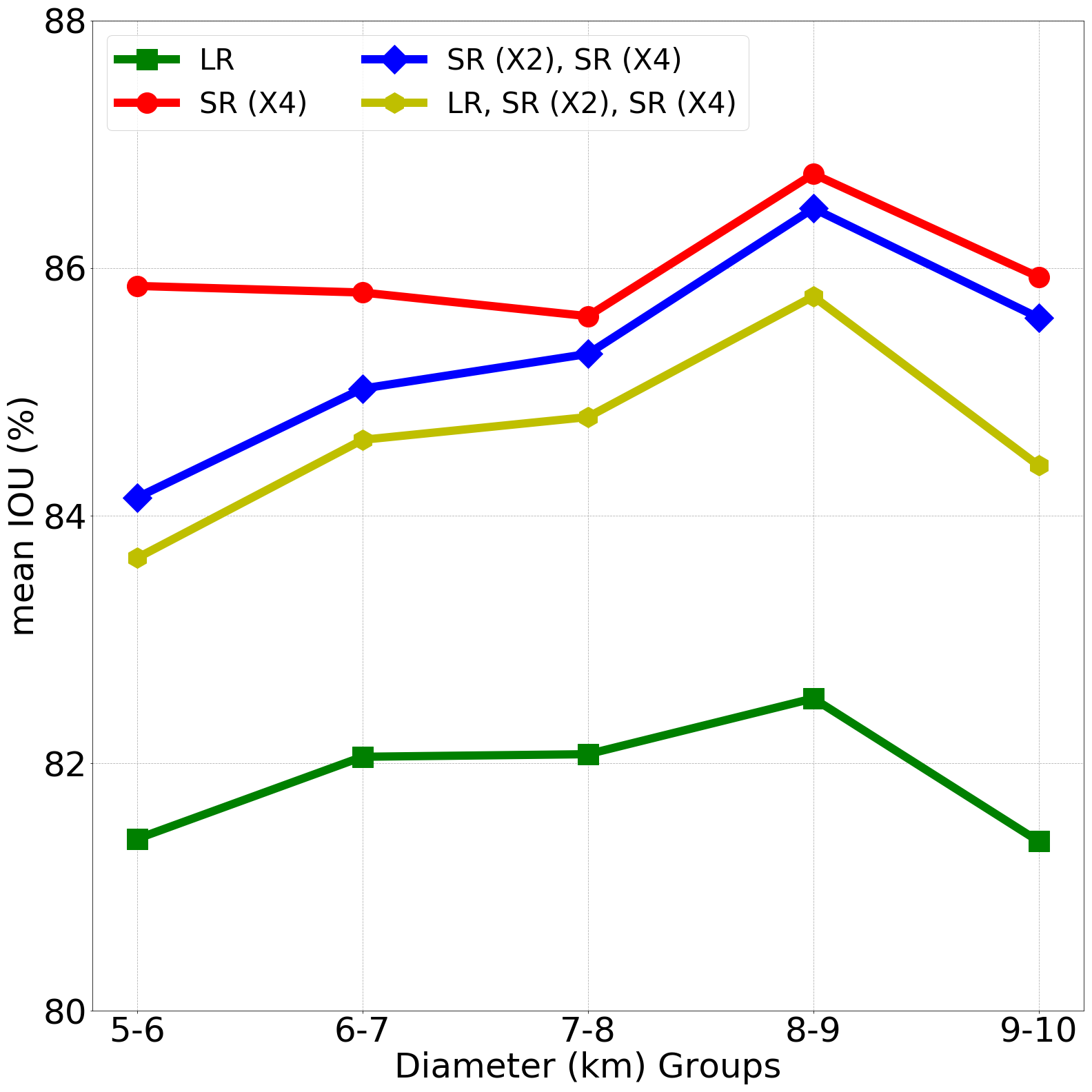}
    \caption{Localization performance in different diameter ranges.}
    \label{fig:loc_perform}
\end{figure}

For fair comparison in localization performance, we have only considered craters detected by all three models: LR, SR ($\times2$), and SR ($\times4$). Figure~\ref{fig:loc_perform} shows the mean localization performance in different diameter ranges. 
In all diameter ranges, the localization performance of SR models is significantly better than the LR model. 
The overall localization performances (mean $\pm$ standered deviation) for a diameter range of $5$ to $10$ km are $81.85\pm9.73$, $85.92\pm8.99$, $85.04\pm9.07$, and $84.48\pm9.33$ for LR, SR ($\times4$), combine outputs of SR ($\times2$) and SR ($\times4$), and  combine outputs of LR, SR ($\times2$) and SR ($\times4$), respectively. 
Also, we have evaluated the localization performance of the combined results of multiple models, as shown in Figure~\ref{fig:loc_perform} (indicated by dashed lines). The localization performance of the combined outputs depends on the models chosen. For example, when we combine the models, SR ($\times2$) and SR ($\times4$), the localization performance falls between SR ($\times2$) and SR ($\times4$)  (black dashed line in Figure~\ref{fig:loc_perform}). Similar trends were observed in other model combinations.
In other words, the localization performance of combined results from multiple models is typically between the localization performances of the individual models. 
The improvement in localization performance in SR models can be due to several factors, such as sharper boundaries, more pixel numbers, and distinguishable craters and non-craters regions in predicted images from SR models. Hence, we concluded that integrating the SR model in the crater detection study improved the localization performance.

\subsubsection{Effectiveness for detecting overlapping craters}
Craters on planetary surfaces are present in diverse sizes and shapes.
Planetary surfaces, such as the Lunar, contain an extensive number of craters, and in some instances, these craters can be overlapped. Investigating these overlapping craters provide an understanding of surface erosion, degradation, and chronological aspect~\cite{wang2021improved}. 
To separate the overlapping craters from other craters, we use the formula provided by Ali-Dib et al.~\cite{ali2020automated} as,

\begin{multline}
    (r_{gt} - r_{pred})^2 < (x_{gt} - x_{pred})^2 + \\(y_{gt} - y_{pred})^2 < (r_{gt} + r_{pred})^2
\end{multline}

Where, ($x_{gt},y_{gt}$) is the center and $r_{gt}$ is the radius of the ground truth crater, and ($x_{pred},y_{pred}$) is the center and $r_{pred}$ is the radius of the predicted crater.

The total overlapping craters of diameter range 5-10 km in the ground truth is 5391.
It is evident from Table~\ref{tab:overlap_sr} that detection results obtained from SR models are better compared to LR. 

The best recall was obtained from the combined detection of predicted images from SR ($\times2$) and SR ($\times4$).

\begin{table}[ht!]
    \centering
    \caption{Impact of SR on detecting overlapping craters.}
    \begin{tabular}{p{0.14\textwidth}p{0.08\textwidth}p{0.13\textwidth}} \hline
        Models & Recall (\%) & Matched Craters \\ \hline
         LR &  47.15 & 2542 \\ 
         SR ($\times2$) & 56.13 & 3026 \\
         SR ($\times4$) & 53.83 & 2902 \\
         SR ($\times2$), SR ($\times4$) & 58.99 & 3180 \\ \hline
    \end{tabular}
    \label{tab:overlap_sr}
\end{table}

\subsubsection{Effectiveness for detecting degraded craters}

Robbins et al.~\cite{robbins2019new} provided an attribute, i.e., ARC\_IMG, to estimate the proportion of the complete rim delineated.
A fresh crater would typically exhibit a higher ARC\_IMG value (closer to 1), whereas an older crater is expected to be eroded and overlapped by other craters; consequently, the lost part of the rims will have a lower ARC\_IMG.
Therefore, as stated in Ali-Dib et al.~\cite{ali2020automated}, the ARC\_IMG can be considered a loose proxy of the degradation quality of the craters.

Following Ali-Dib et al.~\cite{ali2020automated}, we divided the ARC\_IMG parameter into three ranges: $\geq$ 0.95, 0.75 to 0.95, and 0.5 to 0.75, as shown in Table~\ref{tab:arc_img_sr}.
It can be observed that larger ARC\_IMG range (i.e., $\geq$0.95), the recall of all models (i.e., LR, SR ($\times2$), SR ($\times4$)) is higher compared to lower ARC\_IMG ranges.
In lower ARC\_IMG, the recall drops significantly; consequently, detecting potentially degraded craters is challenging.
However, introducing SR models in crater detection study significantly improves recall compared to the LR counterpart. Also, the performance improvement in the lower ARC\_IMG range in the SR models is better than the LR counterpart. For example, in the case of ARC\_IMG $\geq$ 0.95, the performance improvement in SR ($\times2$) compared to LR is 6.14\%, whereas, in ARC\_IMG range 0.5 to 0.75, it is 9.52\%. Therefore, we can conclude that integrating the SR model in the crater detection study will improve the performance for detecting degraded craters.
The combined detection outputs of predicted images from SR models (SR ($\times$2) and SR ($\times$4)) have the highest improvement in all ARC\_IMG ranges. Particularly, in the most degraded craters scenario (in ARC\_IMG range between 0.5 to 0.75), the performance improvement is 15.01\% compared to LR.

\begin{table}[ht!]
    \centering
    \caption{Recall comparison of different models (i.e., LR, SR ($\times2$), SR ($\times4$)) with respect to ARC\_IMG.}    
    \begin{tabular}{p{0.1\textwidth}p{0.07\textwidth}p{0.07\textwidth}p{0.07\textwidth}p{0.07\textwidth}} \hline
         ARC\_IMG & LR & SR ($\times2$) & SR ($\times4$) & SR ($\times2$), SR ($\times4$)\\ \hline
         $\geq$0.95 & 87.27 & 93.41 & 92.67 & 94.69\\ 
         0.75 to 0.95 & 64.91 & 73.61 & 71.87 & 77.76\\
         0.5 to 0.75 & 37.79 & 47.31 & 45.30 & 52.80\\
         \hline
    \end{tabular}
    \label{tab:arc_img_sr}
\end{table}

\subsection{Detection of untrained size range using arbitrary-scale SR images}

We also utilized the advantage of arbitrary scale super-resolution for detecting craters on untrained size ranges. In this particular experiment, we have used a trained crater detection model on LR images, which has a size of $256\times256$. The model was trained for craters of diameter size 5-10 km, equivalent to a pixel range of 12.5-25 pixels. 
Consequently, most craters can be detected in the 12.5-25 pixels range only.
Employing high-resolution images in testing, such as images of size $512\times512$, the pixel range of 12.5-25 pixels on the geographic range translated to a range of 2.5-5 km. This observation implies that we can detect lower-diameter craters by utilizing high-resolution images during testing.

To obtain high-resolution images, we utilize an arbitrary scale super-resolution model that can increase resolution in multiple scales with a single model. 
For this study, we have used arbitrary scale super-resolution with scale factors of 2 and 4. The scale factor 2 increases the image size from $256\times256$ to $512\times512$. 
The predicted images from the SR model with a scale factor of 2 are used during testing to detect craters of size range 2.5-5 km. The total number of crater detected are 48,495. 
We also compare the performance with Robbins et al.~\cite{robbins2019new} catalog with a size range of 2.5-5 km; the precision, recall, and F$_1$-score are 63.52\%, 38.15\%, and 47.64\%, respectively.

\begin{figure}[ht!]
    \centering
    \includegraphics[width=0.5\textwidth,trim=0.2cm 1.3cm 0.1cm 1.3cm, clip]{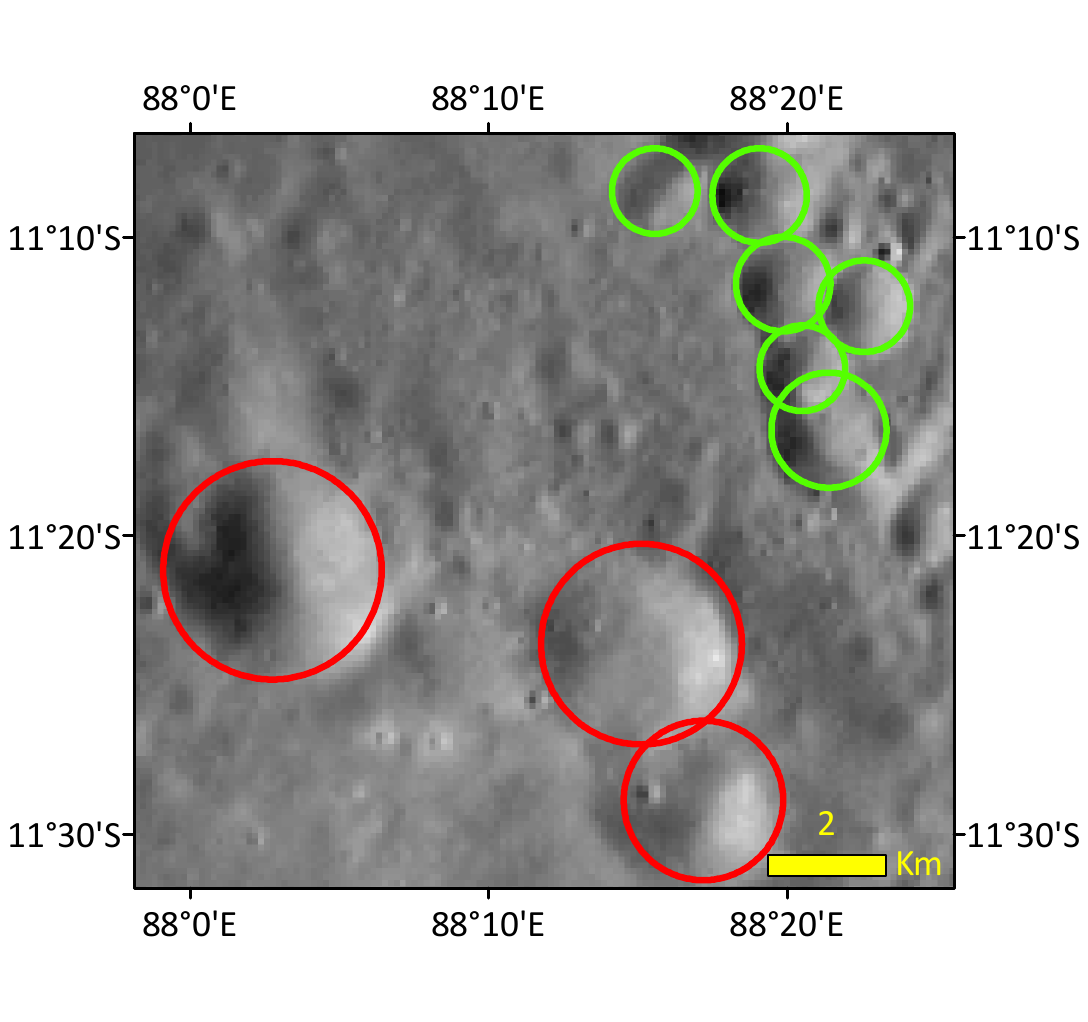}
    \caption{Visual inspection of the detected small size craters. Color coding: Red: detected craters of diameter size of 2.5-5 km, Green: detected craters of diameter size of 1.25-2.5 km.}
    \label{fig:sr_untrained}
\end{figure}

Similarly, for scale factor 4, the image size increased to $1024\times1024$, enabling the detection of craters within the size range of 1.25-2.5 km. In this case, the total number of detections is 1,32,418 craters, and yielding precision, recall, and $F_1$-score are 49.36\%, 27.37\%, and 35.22\%, respectively. Figure~\ref{fig:sr_untrained} visually depicts some of the detected craters using this process; through visual inspection, we can say that the detected craters are true craters.

This analysis highlights that by using arbitrary scale super-resolution, we can detect smaller craters without specific training; however, it is evident that the performance is compromised. This may be because a lesser number of craters are present for training the model compared to the abundance of small-size range craters. Hence, the future direction could be improving the crater detection performance on untrained size ranges.

\section{Conclusion}
\label{sec:Conclusion and Future Work}
We analyzed the impact of super-resolution on crater detection performance across various settings, such as involving overlapping craters and crater shape quality. 
Our findings indicate that the SR approach improves the crater detection performance significantly. 
Notably, it also improves the localization performance of the craters in all diameter ranges.
Additionally, we explored the combination of the outcomes from multiple SR models to increase the detection of the craters.
Finally, we have utilized an arbitrary scale SR model to detect small-size craters without specific training of the crater detection network for small-size craters.
We conclude that SR is the effective solution to improve the resolution of lunar surface images, ultimately leading to improved performance on crater detection.

\section*{Acknowledgment}
This research was partially supported by the ISRO, Department of Space, Government of India, under the Award number ISRO/SSPO/Ch-1/2016-17. 
Atal Tewari is supported by Tata Consultancy Services (TCS) research scholarship.
Any opinions, findings, conclusions, or recommendations expressed in this material are those of the author(s) and do not necessarily reflect the views of the funding agencies. 
We also acknowledge using data from NASA's LRO and SELENE Kaguya spacecraft. 
    
\bibliographystyle{IEEEtran}
\bibliography{paper_bib}

\end{document}